\def\be{\begin{equation}} 
\def\ee{\end{equation}} 
\def\bea{\begin{eqnarray}} 
\def\eea{\end{eqnarray}} 

\newcommand{\tphi}{\tilde{\phi}}

\documentclass{ws-mpla}

\begin{document}

\markboth{Robert H. Brandenberger}
{Alternatives to Cosmological Inflation}

\catchline{}{}{}{}{}

\title{ALTERNATIVES TO THE INFLATIONARY PARADIGM OF STRUCTURE FORMATION}

\author{\footnotesize ROBERT H. BRANDENBERGER}

\address{Physics Department, McGill University, 3600 University Str.,\\ 
Montreal, Quebec, H3A 2T8, Canada\\
and\\
Institute of High Energy Physics, Chinese Academy of Sciences, P.O. Box 918-4\\
Beijing 100049, P.R. China\\
rhb@physics.mcgill.ca}

\maketitle

\pub{Received (Day Month Year)}{Revised (Day Month Year)}

\begin{abstract}
The inflationary paradigm, although very successful phenomenologically, suffers from
several conceptual problems which motivate the search for alternative scenarios
of early universe cosmology. Here, two possible alternatives will be reviewed. -
``string gas cosmology" and the ``matter bounce". Their
successes and problems will be pointed out.

\end{abstract}

\ccode{PACS Nos.: include PACS Nos.}

\section{Introduction}	

The inflationary paradigm of modern cosmology is based on the assumption that
there was a time period starting at time $t_i$ and ending at time $t_R$ during which
the universe undergoes accelerated expansion \cite{Guth}. Inflation explains the
overall homogeneity and isotropy of the universe. Assuming that the universe was
flat enough and large enough initially to reach the phase of accelerated expansion,
inflation can explain the large size and entropy of the current universe, and its
spatial flatness. Most importantly, however, the inflationary paradigm gave rise
to a predictive theory of the origin of the primordial cosmological perturbations
\cite{Mukh} (see also [\refcite{Press,Sato}]).

{F}igure 1 is a space-time sketch of inflationary cosmology. The vertical axis is
time, the horizontal axis indicates physical length. The Hubble radius $H^{-1}$
is the key length scale in cosmology. It separates small scales on which
matter forces dominate from large scales where matter is frozen in and
gravity dominates. The accelerated expansion of space which inflation provides
results in wavelengths which start out at early time with wavelength smaller
than $H^{-1}$ being stretched to become super-Hubble. Thus it is possible to have
a causal generation mechanism for fluctuations. Since any classical matter
present at the beginning of inflation red-shifts, it is reasonable to assume
\cite{Mukh} that perturbations emerge as quantum vacuum fluctuations.
This mechanism produces an almost scale-invariant spectrum of cosmological
perturbations. Thus, the first key requirement on a theory of structure formation
which can explain current data is satisfied.
The perturbations are frozen in at Hubble radius crossing
and propagate on super-Hubble scales without being sourced by
other matter effects. This ``coasting" of the fluctuations on super-Hubble
scales is the second crucial requirement in order to obtain the
characteristic oscillations in the angular power spectrum of the cosmic
microwave anisotropies.
     
\begin{figure}[htbp]
\includegraphics[scale=0.5]{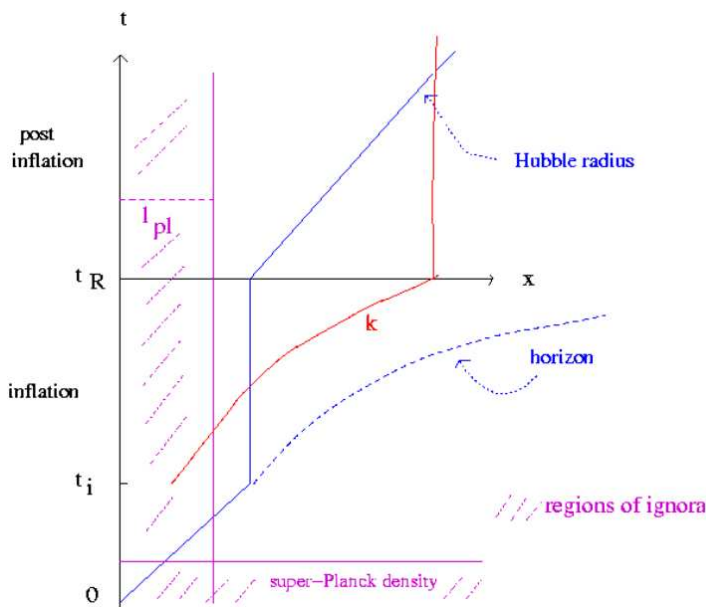}
\caption{Space-time sketch of inflationary cosmology. The vertical axis is time,
the horizontal one is physical distance. The length scales shown are the Hubble
radius (vertical during the inflationary phase - assumed here to be exponential
expansion), the horizon (dashed line) and the physical wavelength corresponding
to a fixed co-moving scale (the curve indicated by the label $k$). This figure also
llustrates the trans-Planckian problem for fluctuations
in inflationary cosmology: If the period of inflation is sufficiently long,
then at the beginning of the inflationary phase the physical wavelength
of all scales which are being observed today is smaller than
the Planck length (the vertical curve indicated by $l_{pl}$). Finally, there
is a high density zone of ignorance which may extend to densities relevant
for inflation.}
\end{figure}


Current models of inflation are based on using general relativity as the
theory of space-time, and making use of a scalar matter field to generate
the phase of accelerated expansion (see, however, [\refcite{Starob}] for the
initial model in which it is correction terms to the gravitational action which
lead to inflation). More specifically, it is required to have a phase during
which the energy-momentum tensor of matter is dominated by the 
potential energy of the scalar field. 

The reliance of inflationary cosmology on scalar fields satisfying
stringent constraints constitutes a first problem for inflation. In order
to make sure that the scalar field rolls sufficiently slowly for a sufficiently
long time, its potential must obey various ``slow-roll" conditions. In
addition, initial conditions for the scalar field must be appropriately
chosen, and the scalar field must be correctly coupled to regular
matter to enable energy transfer from it to regular matter. Once these
conditions are satisfied, the slope of the potential has to be
fine-tuned in order for the amplitude of the induced density
fluctuations not to exceed the observational bounds \cite{Adams}.

A further problem for the inflationary scenario (at least when realized
with scalar matter fields) is the presence of an initial singularity
\cite{Borde}. Thus, inflationary cosmology cannot provide a complete
description of the early universe. 

A more serious problem for inflation is the {\it trans-Planckian problem}
for cosmological fluctuations \cite{RHBrev0,Jerome}. The same
mechanism which inflates the wavelength of fluctuations such that
they begin on sub-Hubble scales leads to the conclusion that,
provided inflation lasted just a few e-foldings of expansion more
than what is required for the scenario to solve the homogeneity
and flatness problems, then the physical wavelength of all scales
which are being observed today is less than the Planck length at
the beginning of inflation. Thus, as illustrated in Figure 1, the
fluctuations emerge from a ``zone of ignorance" where we have
no right to use the Einstein action. It is easy to construct toy models
of trans-Planckian physics which lead to large corrections in
the predictions of inflationary cosmology \cite{Jerome}.

Along the same lines, the applicability of the Einstein effective
field theory equations for the background space-time can be
questioned at energy scales relevant for inflation (about
$10^{16}$ GeV in single field slow-roll inflation models with
simple potentials). All approaches to quantum gravity tell us
that terms in the action other than the Einstein term will dominate
at scales close to the quantum gravity scale. The quantum gravity
scale is close to the energy scale of inflation, and thus using
the Einstein action at these energies may be giving us the
wrong result. These conceptual problems of inflation are
discussed in more detail in [\refcite{RHBrev2}].
  
In light of the conceptual problems outlined above it is important
to study possible alternative scenarios for early universe
cosmology. In the following we will introduce two alternative
scenarios, ``string gas cosmology" and the ``matter bounce".
Both of these scenarios are immune to the trans-Planckian
problem for cosmological fluctuations. Both a designed to
yield non-singular cosmologies. Both make predictions with
which they can be distinguished from inflationary cosmology,
but both also have their set of conceptual problems.
 
\section{String Gas Cosmology}

String gas cosmology \cite{BV} (see also [\refcite{Perlt}] and 
[\refcite{BattWat,SGCreview}] for reviews with extensive lists
of references) is a toy model of the very early universe which
makes use of key new degrees of freedom and symmetries
of string theory (which are not present in quantum particle
theories). It is based on coupling a gas of closed string matter to
a background space-time geometry.

We assume that all spatial sections are compact. For
simplicity, we assume that they are toroidal, with $R$
denoting the radius of the torus). In this case,
the degrees of freedom of closed strings include, in
addition to the momentum modes whose energies
are quantized in units of $1/R$, string winding modes
whose energies are quantized in units of $R$, and
string oscillatory modes (whose energies are independent
of $R$). For comparison, the only degrees of freedom which
point particles have are momentum modes.

The number of oscillatory modes of a closed string increases
exponentially with energy. Hence \cite{Hagedorn} there
is a maximal temperature of a gas of strings, the ``Hagedorn
temperature" $T_H$. The presence of string winding modes
leads to a symmetry of the string mass spectrum under the
transformation $R \rightarrow 1/R$ (in string units). Under
this transition, momentum and winding modes are exchanged.
This is a special case of the more general T-duality symmetry.

Let us imagine a contracting box of strings in thermal equilibrium.
Initially, almost all of the energy is in the momentum modes which
are light at large $R$. Thus, the temperature $T$ will rise as $R$
decreases as in standard particle cosmology. However, as $T$
approaches $T_H$, the energy will gradually drift from the momentum
modes to the oscillatory modes. Once $R$ decreases below $1$,
the energy will move to the winding modes, the modes which are
light at small values of $R$, and the temperature will get smaller
as $R$ decreases further. The temperature-radius evolution is
sketched in Figure 2 \cite{BV}.

\begin{figure}[htbp]
\includegraphics[scale=0.3]{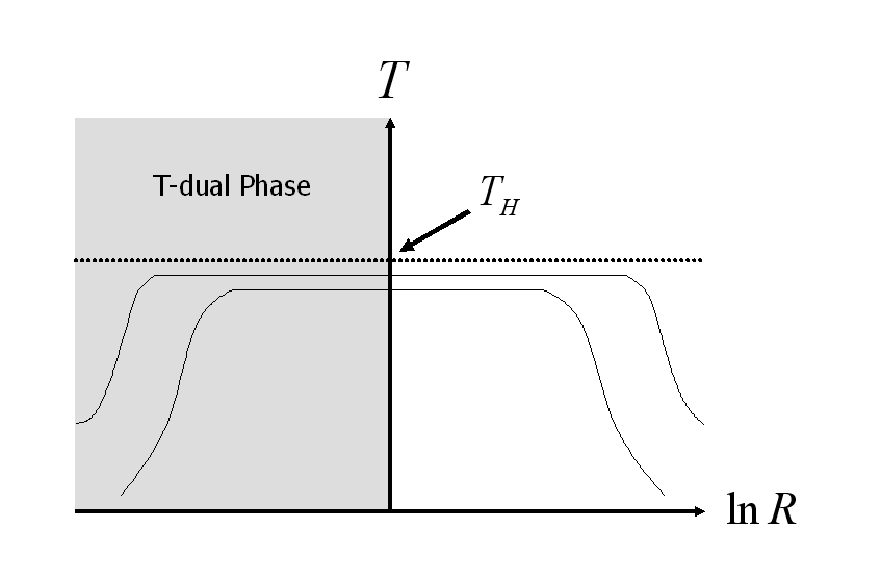}
\caption{Temperature-size relation in string gas cosmology. $R$ is the
radius of the torus, $T$ the temperature. The temperature approaches
but never reaches the limiting temperature $T_H$.}
\end{figure}

In order to determine the space-time dynamics of string gas
cosmology, we need to know how $R$ evolves as a function of
time $t$. There are two possibilities. The first, the one we adopt
here, is that the universe begins close to the Hagedorn
temperature. This configuration is a metastable fixed point
(a fixed point because of the T-duality symmetry, metastable
because the decay of string winding modes into loops
will lead to a dynamical breaking of this symmetry). The
resulting time evolution of the scale factor is sketched in
Figure 3. Another possibility is that $R$ begins much smaller than 
string scale and monotonically increases. This will give
rise to a cosmology in which $T(R)$ evolves as in a bouncing
universe. 

\begin{figure}[htbp]
\includegraphics[scale=0.5]{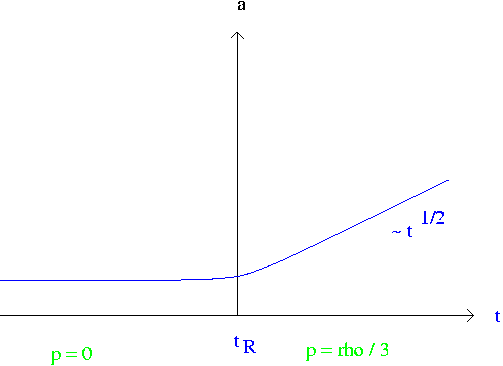}
\caption{The time dependence of the scale factor $a$ in string gas cosmology.}
\end{figure}

As is apparent from the above, there is no temperature singularity
in string gas cosmology \cite{BV}. In addition, one can argue
\cite{BV} that the annihilation of winding modes which is
required in order for space to expand is possible in at most
three spatial dimensions, and that string gas cosmology thus
provides an explanation for the fact that we see only three
large spatial dimensions (concerns about this argument
have been raised in [\refcite{Greene,DFM}]). The interaction of
string winding and momentum modes leads to a dynamical
stabilization of all of the size \cite{Patil,Watson} and 
shape \cite{Edna} moduli of the extra dimensions, without
the need for any additional inputs. This progress has
been reviewed in detail in \refcite{RHBrev3}. The only modulus
field which must be stabilized ``by hand" is the dilaton.
This can be done \cite{DFB}, however, using non-perturbative techniques
of the same type which are used for modulus stabilization in
string inflation model building.

\begin{figure}[htbp]
\includegraphics[scale=0.5]{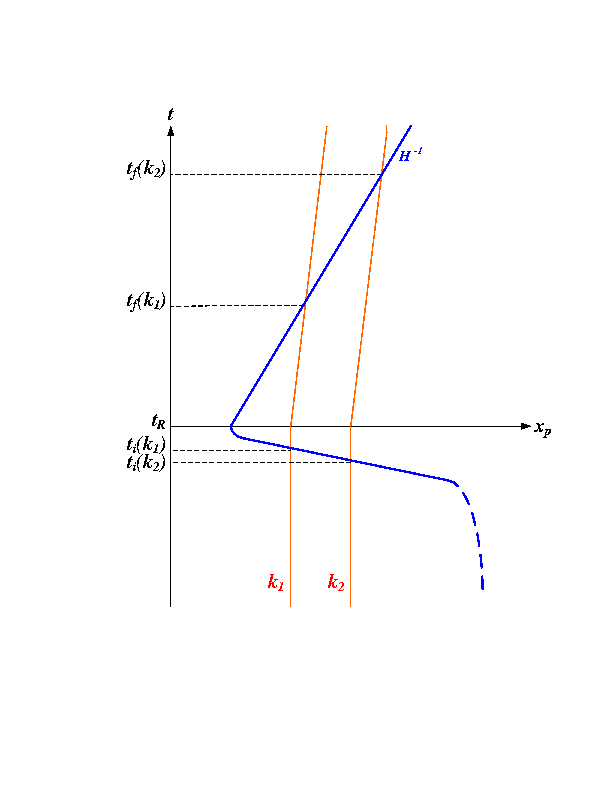}
\caption{Space-time sketch in string gas cosmology. The vertical
axis is time, the horizontal axis physical distance, as in Figure 1. Two
co-moving scales are shown, and the time when a scale $k$ exits
the Hubble radius at the end of the Hagedorn phase is denoted $t_i(k)$.
The time $t_R$ denotes the end of the Hagedorn phase when the
winding modes have annihilated into string radiation.}
\end{figure}

Let us now turn to the issue of fluctuations in string gas
cosmology. Figure 4 presents a space-time sketch of
string gas cosmology. Since the Hagedorn phase
is quasi-static, the Hubble radius tends to infinity
as we get deep into the Hagedorn phase. As is clear
from the sketch, fixed co-moving scales which are
subject of observational cosmology today originate
inside the Hubble radius. If the energy density at
the transition time $t_R$ corresponding to the end of the
Hagedorn phase is of the order of $10^{16}$ GeV, then
the physical wavelength during the Hagedorn phase
corresponding to the present
Hubble radius is about $1$ mm, i.e. a scale in the
far infrared, well removed from the trans-Planckian
zone of ignorance.

Unlike in inflationary cosmology, where the fluctuations
are of quantum vacuum nature because any classical
matter is being red-shifted at an accelerated rate,
in string gas cosmology the dominant fluctuations are
thermal fluctuations of the dominant matter, namely
thermal string fluctuations. As was recently realized
\cite{NBV}, thermal string gas fluctuations in the Hagedorn
phase yield a scale-invariant spectrum of cosmological
perturbations at late times. Thus, string gas cosmology
provides an alternative to the theory of cosmological
inflation in terms of providing a causal theory of structure
formation.

The procedure for computing the spectrum of cosmological
perturbations and gravitational waves is as follows: first,
we compute the matter fluctuations on sub-Hubble scales
during the Hagedorn phase making use of results from
closed string thermodynamics \cite{Deo}. Next, for
each co-moving scale $k$, we compute the resulting metric
fluctuations at the time $t_H(k)$ when the mode exits the
Hubble radius, making use of the Einstein perturbation
constraint equations. Finally, the metric fluctuations evolve
freely on super-Hubble scales until they re-enter the
Hubble radius at late times.

We start by writing the metric including linear cosmological
perturbations and gravitational waves in the following
gauge-fixed form (see e.g. \cite{MFB} for an in-depth
review of the theory of cosmological perturbations, and
\cite{RHBrev1} for a brief overview):
\begin{equation}
d s^2 \, = \, a^2(\eta) \bigl( (1 + 2 \Phi)d\eta^2 - [(1 - 
2 \Phi)\delta_{ij} + h_{ij}]d x^i d x^j\bigr) \,. 
\end{equation}
Here, $\eta$ is conformal time, $\Phi(x,\eta)$ describes the
cosmological perturbations (assuming the absence of
anisotropic stress), and the transverse traceless tensor $h_{ij}$
represents the gravitational waves.

Inserting into the perturbed Einstein equations yields
\begin{equation}  
\langle|\Phi(k)|^2\rangle \, = \, 16 \pi^2 G^2 
k^{-4} \langle\delta T^0{}_0(k) \delta T^0{}_0(k)\rangle \, , 
\end{equation}
and
\begin{equation} 
\label{tensorexp} \langle|h(k)|^2\rangle \, = \, 16 \pi^2 G^2 
k^{-4} \langle\delta T^i{}_j(k) \delta T^i{}_j(k)\rangle \,\, (i \neq j) \, . 
\end{equation}
We thus conclude that the cosmological perturbations are given
by the energy density fluctuations whereas the gravitational
waves are determined by the off-diagonal stress fluctuations.

For thermal matter, the energy density fluctuations are given
by the specific heat capacity $C_V$ (where $V$ indicates the
volume). In a box of radius $R$, one has
\begin{equation}
\langle \delta\rho^2 \rangle \,  = \,  \frac{T^2}{R^6} C_V \, . 
\end{equation}
For a gas of closed strings on a space with stable winding modes,
the specific heat capacity has the holographic form
\begin{equation} 
C_V  \, \approx \, 2 \frac{R^2/\ell_s^3}{T \left(1 - T/T_H\right)}\, . 
\end{equation} 
at temperatures close to the Hagedorn value.

Combining these results, we obtain the following power spectrum of
cosmological perturbations
\begin{eqnarray}
P_{\Phi}(k) \, &\equiv& \,  k^3 |\Phi(k)|^2 \, 
= \, 8 G^2 k^{-1} <|\delta \rho(k)|^2> \nonumber \\ 
&=& \, 8 G^2 k^2 <(\delta M)^2>_R  \, 
= \, 8 G^2 k^{-4} <(\delta \rho)^2>_R \nonumber \\ 
&=& \, 8 G^2 {T \over {\ell_s^3}} {1 \over {1 - T/T_H}} 
\end{eqnarray}
where for each value of $k$, the temperature $T$ is evaluated at the time
$t_H(k)$ of Hubble radius crossing, and $\ell_s$ denotes the string
length.

Since the temperature is approximately constant during the Hagedorn
phase, this power spectrum is scale-invariant. Taking into account the
fact that larger scales exit the Hubble radius at slightly higher temperatures,
one obtains a slight red tilt of the spectrum (more power on longer wavelengths).
These are identical to the results obtained in inflationary models.

The predictions of string gas cosmology can be differentiated from those
of the inflationary scenario by considering the spectrum of
gravitational waves \cite{BNPV1}. By an analysis similar to the one
above, we find the following power spectrum of gravitational waves:
\begin{eqnarray}
P_{h}(k) \, &\equiv& \, k^3 |h(k)|^2 \, 
= \, 16 \pi^2 G^2 k^{-1} <|T_{ij}(k)|^2>  \\ 
&=& \, 16 \pi^2 G^2 k^{-4} <|T_{ij}(R)|^2>  \, 
\sim \, 16 \pi^2 G^2 {T \over {\ell_s^3}} (1 - T/T_H) \nonumber \, .
\end{eqnarray}
The key ingredient from string thermodynamics is 
\cite{Nayeri,BNPV2}
\begin{equation}
<|T_{ij}(R)|^2> \, \sim \, {T \over {l_s^3 R^4}} (1 - T/T_H) \, .
\end{equation}

This spectrum is once again scale invariant, but it has a slight blue
tilt, whereas the gravitational wave spectrum in inflationary
models always has a slight red tilt. The origin of the slight blue
tilt is easy to understand. Long wavelengths exit the Hubble radius
deeper in the Hagedorn phase when the pressure is closer to 
zero and hence the off-diagonal pressure fluctuations are
smaller \cite{BNPV1}.

Let us summarize the requirements which have to be satisfied
in order for the string gas cosmology structure formation scenario
to work. Firstly, there needs to be a quasi-static Hagedorn phase.
Based on the string symmetries we have discussed, we expect
such a phase to exist. Obviously, this phase cannot be described
using Einstein gravity since Einstein gravity does not obey 
the T-duality symmetry of string theory. Dilaton gravity is not
satisfactory, either \cite{Betal}, since it does not include all of the degrees
of freedom required to have S-duality \footnote{Note that the
objections to string gas cosmology of [\refcite{KKLM,KW}] are
based on computations done in the context of dilaton gravity and do
not apply to the setup described here.} At the
moment there is no satisfactory effective field theory description of
the Hagedorn phase of string gas cosmology (for some attempts in
this direction see [\refcite{FKB}]).

The second key requirement for our scenario is that the specific heat
capacity has the holographic scaling $C_V(R) \sim R^2$. Finally,
on infrared scales, the perturbed background equations of motion must
reduce to those in Einstein gravity. All these criteria can be satisfied
in the non-singular bouncing cosmology of [\refcite{Biswas,Biswas2}].

\section{Matter Bounce}

Another way to obtain a scale-invariant spectrum of cosmological
perturbations is from quantum vacuum fluctuations which exit
the Hubble radius in a contracting matter-dominated universe
\cite{Wands,FB2,Wands2,Pinto}. Provided that this phase of
contraction can be connected smoothly to the expanding phase
of standard cosmology, a paradigm alternative to inflation
for generating the observed structure will result.

There are various ways to obtain a non-singular bounce.
One way is by introducing new matter fields with opposite
sign kinetic terms (see e.g. [\refcite{Peter,Fabio,Wands2,quintom}]).
A second way is by making use of specifically chosen higher
derivative terms in the gravitational action. A special
choice of the gravitational action which leads to a non-singular
bounce and at the same time is ghost-free is given in [\refcite{Biswas}]
and discussed in more detail in [\refcite{Biswas2}]. A new
realization of the first approach is the ``Lee-Wick bounce" \cite{LW},
a bouncing cosmology obtained from the scalar sector of the
Lee-Wick Standard Model \cite{LWSM}, an alternative to supersymmetry
for solving the hierarchy problem of the Standard Model of particle
physics.

The scalar sector of the Lee-Wick Standard Model is given by
the following Lagrangian:
\be
{\cal L}  \, =  \, {1 \over 2}  \partial_{\mu} \phi \partial^{\mu} \phi 
- \frac{1}{2} \partial_{\mu} \tphi \partial^{\mu} \tphi 
+ \frac{1}{2} M^2 \tphi^2 
- \frac{1}{2} m^2 \phi^2 - \frac{\lambda}{4} (\phi - \tphi)^4 \, ,
\ee
where $\phi$ is the Higgs and $\tphi$ is its Lee-Wick partner. The
mass scale $M$ is the scale of the new physics. We expect
it to be much larger than the masses of the Standard Model particles.

The above Lagrangian gives rise to a non-singular bounce in the
following way \cite{LW}: We begin the evolution in a contracting phase
with both matter fields oscillating, and the energy dominated
by $\phi$. Thus, the amplitude of $\phi$ is much larger
than that of $\tphi$. The amplitudes of oscillation of both fields scale
as $a^{-3/2}$. Eventually, the amplitude of $\phi$ reaches the value
$m_{pl}$ (Planck mass), and $\phi$ stops to oscillate. Its amplitude
now increases only slowly, whereas the amplitude of $\tphi$ 
continues to increase as $a(t)^{-3/2}$. Thus, the (negative) energy density in $\tphi$
rapidly catches up to the (positive) energy density in $\phi$.
From the Friedmann equations
\bea \label{Heq}
H^2 \, &=& \, \frac{8 \pi G}{3} \bigl[ \frac{1}{2} {\dot \phi}^2 - \frac{1}{2} {\dot \tphi}^2 
 \, +  \frac{1}{2} m^2 \phi^2 - \frac{1}{2} M^2 \tphi^2 + \frac{\lambda}{4} (\phi - \tphi)^4 \bigr] \, ,
\\
{\dot H} \, &=& \,  - 4 \pi G \bigl({\dot \phi}^2 - {\dot \tphi}^2 \bigr) 
\eea
it then follows that $H = 0$ and ${\dot H} > 0$ (the latter since $\tphi$ has
large kinetic energy whereas $\phi$ only has small kinetic energy). Thus,
a non-singular bounce is obtained quite naturally. After the bounce,
the universe soon enters a phase of matter-dominated expansion.

{F}igure 5 is a sketch of the space-time which results in the matter bounce
scenario. The vertical axis is time, the horizontal axis physical distance.
The length scales shown are the Hubble radius as well as the
wavelength corresponding to a fixed co-moving scale $k$.

\begin{figure}[htbp]
\includegraphics[scale=0.3]{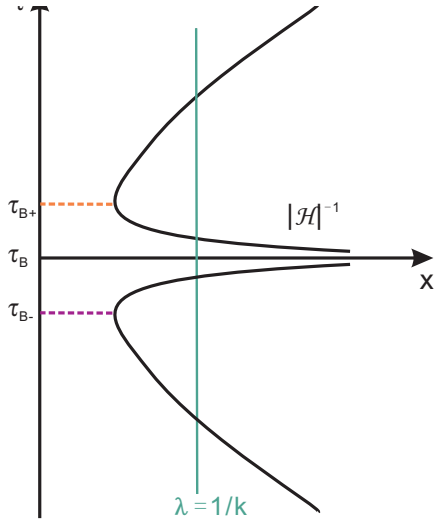}
\caption{Space-time sketch in the matter bounce scenario. The vertical axis
is conformal time $\eta$, the horizontal axis denotes a co-moving space coordinate.
Also, ${\cal H}^{-1}$ denotes the co-moving Hubble radius.}
\end{figure}

Let us now briefly recall how initial quantum vacuum fluctuations in
the contracting matter phase evolve into a scale-invariant spectrum of
cosmological perturbations at late times. It is easiest to work in terms
of the variable $\zeta$, the curvature fluctuation in co-moving coordinates.
The variable $\zeta$ is closely related to the Sasaki-Mukhanov
variable \cite{Sasaki,Mukhanov} $v \, = \, z \zeta$ in terms of which the
action for cosmological perturbations has canonical kinetic
term. The equation of motion for the Fourier mode $v_k$ of $v$ is
\be \label{EOM}
v_k^{''} + \bigl( k^2 - \frac{z^{''}}{z} \bigr) v_k \, = \, 0 \, ,
\ee
where the prime indicates a derivative with respect to conformal time.
If the equation of state of the background is time-independent, then
$z \sim a$ and hence the negative square mass term in (\ref{EOM})
is $H^2$. Thus, on length scales smaller than the Hubble radius,
the solutions of (\ref{EOM}) are oscillating, whereas on larger
scales they are frozen in as standing waves, and their amplitude
depends on the time evolution of $z$.

On super-Hubble scales, the equation of motion (\ref{EOM})
for $v_k$ in a universe which is contracting or expanding as a 
power $p$ of physical time $t$, i.e.
\be
a(t) \, \sim \, t^p \, , 
\ee 
becomes
\be
v^{''}_k \, = \, \frac{p(2p -1)}{(p -1)^2} \eta^{-2} v_k \, ,
\ee
which has solutions
\be
v(\eta) \, \sim \, \eta^{\alpha}
\ee
with
\be \label{alpha}
\alpha \, = \, \frac{1}{2} \pm \nu \,\,\,\, , \,\,\,\, \nu \, = \, \frac{1}{2} \frac{1 - 3p}{1 - p} \, .
\ee
In the case of a matter-dominated contraction we have $\nu = - 3/2$ and
hence
\be
v_k(\eta) \, = \, c_1 \eta^2 + c_2 \eta^{-1} \, ,
\ee
where $c_1$ and $c_2$ are again constants. The $c_1$ mode is the 
mode for which $\zeta$ is constant on super-Hubble scales. However,
in a contracting universe it is the $c_2$ mode which dominates and
leads to a scale-invariant spectrum \cite{Wands,FB2,Wands2}:
\be
P_{\zeta}(k, \eta) \, \sim \, k^3 |v_k(\eta)^2 a^{-2}(\eta) \,
\sim \, k^3 |v_k(\eta_H(k))|^2 \bigl( \frac{\eta_H(k)}{\eta} \bigr)^2 \, 
\sim \, k^{3 - 1 - 2} \, 
\sim \, {\rm const}  \, , 
\ee
making us of the scaling of the dominant mode of $v_k$, the 
Hubble radius crossing condition $\eta_H(k) \sim k^{-1}$, and
the vacuum spectrum $v_k \sim k^{-1/2}$ at Hubble radius crossing.

The above analysis shows that on scales larger than the Hubble
radius, a scale-invariant spectrum of fluctuations in obtained in
the contracting phase. The equation of motion for $\zeta$ has
a singularity at the bounce which is connected with the fact that
the co-moving gauge becomes ill-defined at that point. What was
done in [\refcite{LW}] is therefore to carefully study the evolution
of $\Phi$ through the bounce. The analysis was done numerically
without any approximations, and analytically by solving
the equation of motion for $\Phi$ approximately in the matter-dominated
contracting phase, in the bounce phase, where the Hubble constant
can be modeled as $H(t) = \alpha t$, $t = 0$ denoting the bounce 
time, and finally in the matter-dominated phase of expansion, and
making use of the matching conditions for fluctuations \cite{Hwang,Deruelle}
at the two phase transitions. Note that the use of these matching
conditions is justified since at each transition surface the background 
also satisfies the background matching conditions.

Like string gas cosmology, the matter bounce scenario can solve
the singularity problem of inflation. Again like in string gas cosmology,
the wavelength of fluctuations of current interest in observational
cosmology is in the far infrared at all times, and thus the trans-Planckian
problem for fluctuations does not arise.

Matter bounce scenarios are plagued by the problem of ghosts \cite{Jeon}.
This is a particularly serious problem in models like the Lee-Wick bounce
where new matter fields with phantom behavior are introduced (whereas
the higher derivative gravity model of [\refcite{Biswas}] is free of ghosts).
As long as the energy density remains smaller than those corresponding to
the phantom field mass scale $M$ (as it does in the case of the
Lee-Wick bounce) , the model may be considered as
a low-energy effective field theory emerging from a fundamental theory
free of ghosts. 

A matter bounce scenario is unstable to the presence of radiation
and, more importantly, of anisotropies 
This is a further problem which must be addressed.

\section{Discussion and Conclusions}

The main message of this talk is that there are scenarios other than
inflation which can successfully connect fundamental physics with observations.
The focus of this lecture has been on string gas cosmology and on the
matter bounce scenario. This is, however, by no means a complete list of
alternatives. Pre-Big-Bang cosmology and the Ekpyrotic/Cyclic scenarios
are other alternatives, which for lack of space cannot be described here.

\section*{Acknowledgments}

I wish to thank the organizers of CosPA08 for the invitation
to speak and for their wonderful hospitality in Pohang. I
wish to thank Yifu Cai, Taotao Qiu and Xinmin Zhang for
introducing me to the Lee-Wick model and for collaborating
on [\refcite{LW}], and Yifu Cai, Sugumi Kanno and Subodh Patil
for allowing me to use a figure they each drew. I am grateful to Professor X. Zhang and
the Theory Division of the Institute of High Energy Physics (IHEP) for
their hospitality and financial support. My research is
also supported by an NSERC Discovery Grant and by the Canada Research
Chairs Program. 


\end{document}